# Systematic Mapping Protocol

*The impact of using software patterns during requirements engineering activities in real-world settings.*

José L. Barros-Justo[a]    Ania L. Cravero-Leal[b]    Fabiane B. V. Benitti[c]    Rafael Capilla-Sevilla[d]

[a] School of Informatics (ESEI), University of Vigo, 32004 Ourense, Spain.
[b] Departamento de Ciencias de la Computación e Informática, Centro de Estudios en Ingeniería de Software, Universidad de La Frontera, Temuco, Chile.
[c] Universidade Federal de Santa Catarina (UFSC), Florianópolis, Santa Catarina, Brazil.
[d] Rey Juan Carlos University, Madrid, Spain.

# 1. Introduction

There are many published works, both in Journals and Conferences, about software patterns (mainly in the phases of analysis and design, i.e. architectural and design patterns) and the supposed benefits they offer, but little evidence (real data) in the context of industrial applications has been offered by these publications. We are interested in answer the big question: Are these claims about the benefits of using patterns in industrial contexts real or just a myth? Besides, we will focus our work in the context of requirements engineering (RE), orienting our question to: What is the impact of using software patterns during the RE phase (early activities of the SDLC)? Really worth it? Could the use of software patterns be counterproductive in specific context of software development? Which are the most used software patterns? By which RE activities?

This document details the planning phase of a Systematic Mapping Study (SMS). Our goal is to identify the software patterns used during the RE phase, in real-world setting (i.e., in real projects), not in academia (toy projects) and, to understand the impact of their application, in terms of different characteristics, pertaining to the development process as well as the final product. Through a review of the literature published until January 2017, we will investigate what the research community has reported on the application of patterns in the industrial context, the data supporting these claims, the specific patterns employed, the RE activities, the results (positive or negative) and the metrics used to validate these results.

As a research method we have choose a systematic mapping study due to its adequacy for the exploration of a research area in a systematic way. Following the advice of numerous well-known authors in the area of Evidence-Based Software Engineering (EBSE), we have decided to develop the protocol for the study as an independent document, containing all the details needed to replicate our work by any other researchers, assessing the validity of the study.

The next section presents detailed information about the planning phase of the systematic mapping.

> *"If you don't know where you're going, any road will get you there."*
>
> Lewis Carroll (from Alice in Wonderland)

# 2. Planning

## 2.1. Research goal and the set of research and publication questions

The objective of this systematic mapping study (SMS) and therefore, its main research goal (MRG) is:

- *Identify and classify the patterns that have been used during the requirements engineering phase of software development and assess their impact in real-world settings.*

Considering that the interest of the study is the use of patterns in software development, during the requirements phase, we will use the term "pattern" as a synonym of "software pattern", and always within the context of software engineering.

The main research question (MRQ) for the mapping study is formulated as:

- *What impact has been reported about using patterns during the requirements engineering phase of software development in real-world settings?*

This MRQ is broad enough to allow us to slice the research space in two dimensions: topic research space and publication space. The topic research space is mutable, it depends strongly on the research topic and, therefore the research questions (RQs) are subject to change in every systematic review. On the other hand, the questions in the publication space (Publication Questions, PQs from now on) can be fixed beforehand, as they are very similar in many systematic reviews (Kitchenham B.; Chartes, et al., 2007) (Kitchenham, et al., 2012) (Petersen, et al., 2015).

### 2.1.1. Topic research space

We start by splitting the MRQ in three main categories:

- RQ1: Which patterns have been reported?
- RQ2: Which RE activities have been reported?
- RQ3: Which properties of the software development process, or of the final product, was affected (impacted) by the use of patterns?

And two additional research questions to assess, to a certain extent, the quality of the report itself.

- RQ4: Was the impact of using patterns measured? (Y/N), if Yes, report the metrics used and the amount of the impact in adequate magnitudes (money, time, number of bugs, etc.)
- RQ5: Which was the research method reported?

Table 1 shows the set of the research questions and a brief description of each one.

*Table 1    Description of the Research Questions*

| RQ | Description |
|---|---|
| RQ1: Which patterns have been reported? | Name of the pattern, (verbatim) as it appears in the original source. |
| RQ2: Which RE activities have been reported? | The name of the activity, as it appears in the original source, will be mapped to the standard ISO/IEC/IEEE 29148:2011 (ISO/29148, 2011). An option "Other" exists in case the reported activity can not be mapped to the standard.<br>SR2.1 Elicit requirements and Analyze System Context<br>SR2.2 Verify Stakeholder Requirements with PM<br>SR2.3 Validate Stakeholder Requirements<br>SR2.4 Review System Requirements & External Interfaces<br>SR2.5 Define/Update Traceability between Requirements<br>SR2.6 Verify & Obtain Work Team Approval<br>SR2.7 Validate that System Specs Satisfies Stakeholder Specs |
| RQ3: Which property of the software development process, or of the final product, was affected (impacted) by the use of patterns? | The name of the property as it appears in the original source. If it is possible we will map the properties to some standards, for example, standard ISO/IEC 25010:2011 (ISO/25010, 2011) will be used for product quality features. |
| RQ4: Was the impact of using patterns measured? (Y/N), if Yes, how much was the impact? | If the impact was measured, then report: a) the metrics and b) the amount. For example: bug reduction, 23%; or dollars, 150,000; or development time reduction, 15%... |
| RQ5: Which was the research type reported? | The six types of research proposed by Wieiringa (Wieringa, et al., 2006), see Figure 5 in Data extraction protocol. |

### 2.1.2. Publication space

PQ1: Top venues: Journals/Conferences or Workshops, and their evolution (published works/year)

PQ2: Publications per year (its evolution, no matter the venue)

PQ3: Top-cited papers (normalized number of citations)

PQ4: Active researchers (number of works in the set of selected papers)

PQ5: Researcher's affiliation (Academia/Industry/Both)

## 2.2. Search strategies protocol

*Table 2 Search for papers*

| input | Automated strategy: databases + search string<br>Snowballing: set of initial papers (seeds)<br>Manual: list of Journals, list of Conferences, list of Authors |
|---|---|
| output | set of retrieved works (for Paper Selection phase) |
| participants | r1, r2, r3, r4, r5 and r6 |

*Table 3 Search strategies: Role/Activities*

| participant | activity |
|---|---|
| r1 | Management: assignments, control & validation, integration of results.<br>Monitor: database selection, the construction of the search string, selection of journals, conferences and main authors. |
| r2 & r4 | Automated search strategy (r2: two databases, r4: the other two databases) |
| r3 & r5 | snowballing search strategy (r3: backward, r5: forward) |
| r6 | manual search strategies (journals, conferences, main authors) |

We were interested in finding all the available evidence for two different sections of the mapping study: Related works section (previous systematic reviews or mappings) and the Method section (primary works to extract data and analyse it). The protocol produced two different search strategies, one for each of these two sections.

### 2.2.1. Search Strategies for Related work section:

**Automated search**

The first step was the selection of the online resources to be used. As many others systematic reviews suggest (Santos, et al., 2015) (Wohlin, et al., 2013) (Chen, et al., 2010) (Meho & Sugimoto, 2009), we decided to use four different electronic resources:
- SCOPUS[1] (Indexing service, including: IEEE, ACM & Elsevier, Wiley & Springer)
- Web of Science[2] (Indexing service, including: IEEE, ACM & Elsevier)

---

[1] https://www.scopus.com/home.uri

[2] https://apps.webofknowledge.com

- IEEE Xplore[3] (publisher database: IEEE Digital Library)
- ACM DL[4] (publisher database: ACM Digital Library)

Note the overlapping between the databases (IEEE and ACM publications are covered (indexed) by three of the four resources). This overlapping allows us to reduce the risk of "not finding" some work of interest (theoretical validity).

### Search string creation (and evolution).

We define a general, broad search string, which include the key term pattern, and complement it with terms related to secondary studies: systematic, review and mapping. These key terms came from three different sources:
- Our goal and the set of research questions (RQs),
- Renown publications in the area of software patterns and requirements engineering,
- Some frequently used synonyms.

The output from these three sources was a set composed by the key terms: pattern, patterns, templates, systematic review, literature review, systematic mapping and mapping study.

We add the terms "*software engineering*" and "*requirements engineering*" to add context to the search string. After running some pilot searches, we eliminated some terms from the search string, providing they do not reduce the previous set of retrieved works. Finally, we arrived to a general search string and tailored it to each of the online databases and digital libraries. For example, for SCOPUS the final search string was composed as follows:

**TITLE (pattern\*) AND KEY (systematic OR review OR mapping) AND KEY ((requirements engineering) OR (software engineering))**

We assume that the term "*pattern*" was so important that it should appear in the title of the work, while the rest of the terms are intended to: a) select secondary studies and, b) set the focus in the *requirements engineering* phase or in *software engineering* (broader area).

We ran searches[5] in ACM DL, IEEE Xplore, SCOPUS and WoS, getting 168 works (including 26 duplicates). The sequence to eliminate duplicates was:

1. ACM DL: as this digital library do not retrieve the Abstracts
2. WoS: as it indexed the same content as SCOPUS
3. IEEE Xplore: as it indexed the same content as SCOPUS

---

[3] http://ieeexplore.ieee.org/Xplore/home.jsp
[4] http://dl.acm.org/
[5] on 2016/11/15

*Table 4 Automated search output*

|  | ACM DL | WoS | IEEE Xplore | SCOPUS | Totals |
|---|---|---|---|---|---|
| Retrieved | 4 | 77 | 41 | 46 | 168 |
| Duplicates | 3 | 15 | 8 | --- | 26 |

**Snowballing.**

After applying exclusion criteria (see Papers selection section) we selected a work (Franch, 2015) which is not a real secondary study but a technical briefing and, applied a snowballing process as suggested in (Wohlin, 2014). This work was of interest because it reports on some other studies related with our area of research (patterns and requirements engineering). Using this work as a seed we performed a backward snowballing (the work does not have any citations yet) by analyzing its 16 references. From that list we did not select any work as the exclusion criteria excluded all of them. We use the SCOPUS option "*View all related documents based on references*" and added a second work (Da Silva & Benitti, 2011) and used it as a seed for a new round of the snowballing process. In order to reduce the number of works retrieved by the "*View all related documents based on references*" option we impose some criteria on the results, as follows (SCOPUS syntax):

**Refined to [((((systematic)) AND (literature OR mapping)) AND (pattern\*) AND (LIMIT-TO (DOCTYPE, "cp") OR LIMIT-TO (DOCTYPE, "ar") OR LIMIT-TO (DOCTYPE, "re")) AND (LIMIT-TO (SUBJAREA, "COMP")) AND (LIMIT-TO (EXACTKEYWORD, "Requirements Engineering")) AND (LIMIT-TO (SRCTYPE, "p") OR LIMIT-TO (SRCTYPE, "j"))]**

Where:
- DOCTYPE "cp", "ar" and "re": Conference paper, Article (Journal) and Review (respectively)
- SUBJAREA "COMP" limits the area of interest to Computer Science
- EXACTKEYWORD forces the terms "Requirements Engineering" to be present in the index keywords (author or publisher index terms)
- SCRTYPE limits the *source* to be "p" (Proceedings) or "j" (Journal)

*Table 5 Snowballing search output*

| Source | References | Citations | Related documents | Totals |
|---|---|---|---|---|
| (Franch, 2015) | 16 | --- | 22 | 38 |
| (Da Silva & Benitti, 2011) | 18 | 1 | 35 | 54 |

These 92 works were pass to the next phase: Paper selection.

### 2.2.2. Search Strategies for Primary works:

Four different search strategies will be used in order to fulfill the claims (and warnings) of many reports and guidelines on systematic literature reviews (Badampudi, et al., 2015) (Boell & Cecez-Kecmanovic, 2015) (Kitchenham & Brereton, 2013) and systematic mappings (Garousi, et al., 2016) (Petersen, et al., 2015) (Wohlin, et al., 2013), i.e., to find ALL the available evidence. We will carry on:

- Automated search on four different online databases,
- Snowballing (backward and forward) with works from our Related Work section,
- Manual search of:
  - A set of selected journals and conferences that usually publish research in the areas of interest (software engineering, software patterns, requirements engineering). Two reviewers (r2 & r4) construct independent lists of potential Journals and Conferences and then a third reviewer (r1) merge these lists to produce the final set to be searched. As an example, the annex I shows the final lists produced for this study, both, journals and conferences.
  - Publications from a set of selected authors (rank by productivity, number of published papers and number of citations of works included in our previous set of selected primary works, from previous search strategies). We will visit their personal web pages and/or retrieving their publications profile from DBLP[6].

**Automated search:**

The Automated search was carried out using the same databases as in the Related work section, and due to the same reasons mentioned there.

**Search string creation (and evolution)**

We tried to applied the PICOC Structure suggested by (Kitchenham B.; Chartes, et al., 2007) to identify keywords that could be used to build up the search string. Our first attempts lead to a term assignation as follows:

**Population**: Published scientific literature reporting Software Patterns and/or Requirements Engineering studies.

**Intervention**: studies involving the use of patterns during the requirements engineering phase.

**Comparator**: not applicable.

**Outcomes**: Impact of using patterns (on the final product and/or the development process).

**Context**: Industry (real-world settings).

---

[6] dblp computer science bibliography: http://dblp.uni-trier.de/

Unfortunately, this structure did not offer enough information to extract the key terms for the search string. We discard this strategy and use the following three, complementary ones.

- Deriving major search terms from our goal and research questions: Our goal was established as: *Identify and classify the patterns that have been used during the requirements engineering phase of software development and assess their impact in real-world settings*. From this goal, we extracted: *patterns, requirements engineering, software development and real-world setting*. The first three key terms also came from the RQ1, RQ2 and RQ3, respectively. Finally, our fourth research question (RQ4) contributes with: *impact, intervention, outcome*.
- Conduct pilot testing: We ran a pilot test, searching on the four databases mentioned before with the terms from the previous section. Our goal was to identify other relevant terms, synonyms and alternative spellings that are frequently used in published literature, both by authors and editors/publishers. The Table below shows our findings:

*Table 6   Relevant key terms from pilot search.*

| Source | Keywords |
|---|---|
| (Alebrahim, et al., 2015) | architectural patterns, quality requirements |
| (Bunke, 2015) | security patterns |
| (da Silva, et al., 2015) | requirements patterns, use cases, requirements specification |
| (Ketabchi, et al., 2011), (Mahendra & Ghazarian, 2014), (Franch, 2015) | software requirements patterns |
| (Kolp, et al., 2003) | organizational patterns |

The original set of key terms extracted was: requirements patterns, organizational patterns, security patterns and requirements specification.

We combined the terms extracted from the first strategy and this second one and, produce a simpler set, which encompass all of them: pattern, requirement, engineering, software.

- Connect the resulting terms using Boolean operators to construct the Search String. The final version of the search string (using the syntax for SCOPUS) follows:

**TITLE (pattern\*) AND (engineering AND (software OR requirement))**

### Digital Libraries and Databases.

We use the following electronic resources: ACM DL, IEEE Xplore, SCOPUS and WoS. The search string was tailored to these databases as detailed in the following table:

*Table 7 Search strings*

| Database | Search string | Results |
|---|---|---|
| ACM DL | "query": { {acmdlTitle:(pattern*) AND keywords.author.keyword:(engineering*) AND (keywords.author.keyword:(software) OR keywords.author.keyword:(requirement)) } } | 69[7] |
| IEEE Xplore | "Document Title":pattern* AND "Author Keywords":engineering AND ("Author Keywords":software OR "Author Keywords":requirement*) | 110 |
| SCOPUS | TITLE(pattern*) AND AUTHKEY("engineering") AND AUTHKEY(software OR requirement*) | 302 |
| Web of Science | TI=pattern* AND TS=engineering AND TS=(software OR requirement*) | 803[8] |

### Snowballing search

Snowballing was conducted according to the guidelines by (Wohlin, 2014). The number of Citations as well as the search for the referenced works was performed using SCOPUS.

### Initial set selection

The initial set for the snowballing process will consist of articles from the set of selected works (after completing the selection phase). We will choose the 5 works with the highest number of citations to build up the initial set (seeds for the snowballing process).

### Forward & Backward search

For backward snowballing, we will review the works listed in the reference section of each seed work, limiting the search by date (works dated before 1991 will not be considered, as the first appearance of the term pattern in the context of software engineering occurs around 1992).

### Manual search:

### Conferences and Journals:

Two reviewers (r2 & r3) construct independent lists of potential Conferences, Workshops and Journals and then a third reviewer (r1) merge these lists.

---

[7] These results correspond to searches conducted on 2016/12/06
[8] Web of Science do not have a field author_keywords. The field TS includes: Title, Abstract and Keywords.

**Most cited Authors:**

This is the last strategy, as it depends on previous results and output from the paper's selection process. We will use the author's personal web page or DBLP profile to collect works of potential interest. We will rank, by number of citations, all the studies included in our set of selected works. After that, we will select the first author of the first five top cited studies, in two different categories:

- Five from Academia: although we are interested in real-world settings only, some works reporting industrial experiences were written by academics. That is why we consider them here.
- Five from Industry: the affiliation of the main author at the time of writing the paper.

### 2.2.3. Validate the search

The search process should be validated using a set of known papers provided by an expert in the research area. That set can also be used to build a Quasi-Gold standard (Kitchenham, et al., 2015) (Zhang, et al., 2011) to check the completeness of our automated search strategy, by computing sensitivity (recall). A sensitivity greater than 80% should be considered as a very good value.

A completeness check should be conducted by using DBLP database (test the output from our most cited Authors search strategy and compare with the records in DBLP).

## 2.3. Papers selection protocol

This section offers a detailed guide on how to conduct the selection of secondary works (for the Related work section) and primary works (for the SMS itself).

**Paper Selection phase**

*Table 8   Selection protocol*

| input | works retrieved by search strategies |
|---|---|
| output | set of selected works (for Data extraction phase) |
| participants | r1, r2, r3, r4, r5 and r6 |

*Table 9    Roles and activities for the selection of primary papers*

| participant | activity |
|---|---|
| r1 | Management: assignments, control & validation, integration of results. Establish inclusion/exclusion criteria |
| r2 & r4 | apply selection criteria to works retrieved by snowballing search strategy |
| r3 & r5 | apply selection criteria to works retrieved by automatic search strategy |
| r6 | apply selection criteria to works retrieved by manual search strategies |

### 2.3.1. Paper selection for Related work section

**Input:** works retrieved by the search for related work.

**Exclusion criteria**:

To select the works for the related work section we established the following exclusion criteria (the sequence should be done in this order):

1. Remove duplicates (works indexed by more than one database)
2. Remove works that are not secondary studies (systematic literature reviews or mapping studies)
3. Remove works not focused on patterns and requirements engineering

**Exclusion of works from the automatic search strategy**:

To eliminate duplicates we start by removing duplicate entries in ACM DL as this digital library offers less information than the others (no Abstract information is available), then we removed works retrieved by WoS since IEEEXplore offers the same information and it is more easy to use. We finally excluded duplicate works from IEEEXplore (as they were also indexed by SCOPUS). The Table below summarizes this process and shows the data obtained:

*Table 10   Duplicate works*

|  | SCOPUS | ACM DL | WoS | IEEEXplore | Total |
|---|---|---|---|---|---|
| SCOPUS | --- | 3 | 10 | 8 | 21 |
| IEEEXplore | --- | --- | 5 | --- | 5 |

The uniques works retrieved from each online database are shown in the Table below:

*Table 11  Unique works retrieved by each source*

| SCOPUS | IEEEXplore | WoS | ACM DL | Total |
|--------|------------|-----|--------|-------|
| 46 | 33 | 62 | 1 | 142 |

After removing the duplicates, we discarded the works that were not secondary studies, i.e., systematic literature reviews or systematic mappings. Six works were classified as secondary studies, but five of them were rejected because they were not focused on patterns and requirements engineering.

*Table 12  Sources of secondary works*

| SCOPUS | WoS |
|--------|-----|
| 5 | 1 |

*Table 13  Works not focused on the research area*

| SCOPUS | WoS |
|--------|-----|
| 4 | 1 |

The final output from this phase was only one work (Franch, 2015), so we use it as a seed for the next search strategy, the snowballing (Wohlin, 2014).

**Exclusion of works from the Snowballing search strategy**:

After conducting backward snowballing from the seed paper of Franch (Franch, 2015) (the paper has no citations) we get the following results:

*Table 14  Snowballing output*

|  | References | Citations | Related documents | Total |
|--|------------|-----------|-------------------|-------|
| Snowballing | 34 | --- | 57 | 92 |

After applying the exclusion criteria to 92 works retrieved by the snowballing process we finally selected two works: (Franch, 2015) and (Da Silva & Benitti, 2011).

### 2.3.2. Primary papers selection (for the Research Method section)

**Input**: works retrieved by the search strategies for primary works.

**Exclusion criteria**:
To select the works for the Method section (the SMS itself) we established the following exclusion criteria (the sequence should be done in this order):

Step 0: Duplicate papers were discarded

Step 1: the results from the search strategies were screened on: title, keywords and source (Journal or Conference).
- 1.1. Not in English language.
- 1.2. Not peer-reviewed publications (only accept Journals & Conferences).
- 1.3. None major search terms appear in Keywords (from author or editors).
- 1.4. Books, editorials, tutorials, panels, poster sessions, prefaces, opinions, letters, slide presentations, technical reports and so (grey literature).
- 1.5. Papers published before January 1991 or after December 2016.

Step 2: The resultant papers from Step 1 were evaluated on abstracts only to exclude the studies that were:
- 2.1. Totally irrelevant papers that were retrieved due to poor execution of search string by online search engines (paper's focus is not on the research topic).
- 2.2. Not from the domain of Requirements Engineering (or Software Engineering).
- 2.3. Publications that are unrelated to software patterns application (real use)
- 2.4. Publications that are unrelated to a real-world setting (industry domain)
- 2.5. Pure opinion papers or experience reports (without real data), or surveys where participants provide their opinion (without actual supporting data).
- 2.6. PhD or Master theses were also excluded because relevant publications resulting from the research covered by the theses should have been published in peer-reviewed journals or conferences, so they are available and should been already retrieved and included in this study.

Step 3:
- 3.1. For papers having conference and journal versions, select the most detailed/recent one (first: number of pages; if equal then: most recent).
- 3.2. Full text is not available (reject abstracts, extended abstracts, introduction papers, etc.)

Papers not excluded after the application of the above criteria will be considered as *selected primary works*, and will feed the next process (Data extraction).

Different reviewers, to avoid bias and guarantee the validity of the study, should apply these steps. For the set of works retrieved by every search strategy, we assigned a different pool of reviewers (note the "inversion" of reviewers with respect to searches for the selection of primary works).

*Table 15  Role assignment for paper's selection activities*

| Activity (Search Phase) | Reviewers | Activity (Paper selection Phase) |
|---|---|---|
| automatic search strategy | r2 & r4 | r3 & r5 |
| snowballing search strategy | r3 & r5 | r2 & r4 |
| manual search strategies | r6 | r1 |

To deal with disagreements we applied the inclusive criteria A+B+C+D proposed by (Petersen 2015). Papers classified as "E" were considered borderline and they will be listed in an Appendix of the SMS. We excluded a paper only when both reviewers agreed (category "F") or considered the paper as borderline (category "E").

*Table 16  Dealing with disagreements*

|  |  | Reviewer X | | |
|---|---|---|---|---|
|  |  | Include | Uncertain | Exclude |
| Reviewer Y | Include | A | B | D |
|  | Uncertain | B | C | E |
|  | Exclude | D | E | F |

As a validation of the selection process, we will compute the Kappa statistic between reviewers (that is the reason why we forced the selection process to be conducted by two reviewers independently, as a blind review process).

## 2.4. Data extraction protocol

In the data extraction step the researchers will read the full text of each article identified for inclusion in the review and extract the pertinent data using a standardized data extraction/coding form. This section offers a detailed guide on how to perform the extraction process.

The work will be divided in two sections - RQs & PQs

1. For RQs: divide the set in two halves, like the most common strategy identified by Petersen (Petersen, et al., 2015).
    1.1. First half: This first half will be assigned to reviewers r1 and r2, without any of them know about the other (blind assignment).
    1.2. Second half: This second half will be assigned to reviewers r3 and r4, also as a blind assignment.
2. For PQs: These issues will be assigned to r5 because they are composed only of objective items and, therefore, we think they are much easier to extract.

Create a Data Extraction Form (DEF) in spreadsheet format, with columns for every RQ (selected papers will be the rows). Every single cell should contain:

1. Data extracted
2. A comment indicating: # of page & original text (short one) (see Figure 1)

The reviewers will fill the DEF independently. If conflicts arise then a consensus meeting is held and, if they do not reach an agreement, then a third reviewer will decide.

|   | A | B | C |  |
|---|---|---|---|---|
| 1 |   |   | Research Questions |   |
| 2 | Paper ID | RQ1 (Pattern name) | RQ2 (RE activities) | RQ3 |
| 3 | SP1 |   | SR2.1 Elicit requirements and Analyze System Context |   |
| 4 | SP2 |   | SR2.2 Verify Stakeholder Requirements with PM |   |
| 5 | SP3 | Pattern1 | SR2.3 Validate Stakeholder Requirements |   |
|   |   |   | SR2.4 Review System Requirements & External Interfaces |   |
| 6 |   | Pattern2 | SR2.5 Define/Update Traceability between Requirements |   |
|   |   |   | SR2.6 Verify & Obtain Work Team Approval |   |
| 7 | SP4 |   | SR2.7 Validate that System Specs Satisfies Stakeholder Specs |   |
| 8 | SP5 |   | Other |   |
|   |   |   | Not reported |   |
| 9 | SP6 |   |   |   |
| 10 | … |   |   |   |

Figure 1  Data Extraction Form (DEF), part I

|   | D | E | F |
|---|---|---|---|
| **Research Questions** | | | |
| | RQ3 (Feature impacted) | RQ4 (Metric and amount) | RQ5 (Research type) |
| | | | Evaluation research |
| | | | Solution proposal |
| | | | Validation research |
| | | | Philosophical papers |
| | | | Opinion papers |
| | | | Experience papers |
| | | | Other |
| | | | Not reported |

Figure 2  Data Extraction Form (DEF), part II

Note that when a paper presents more than one pattern, we report the data of each one in a specific line (see Figure 1, line for SP3).

For questions in the Publication Space the template was configured as follows:

|   | A | B | C | D | E | F |
|---|---|---|---|---|---|---|
| 1 | | **Publication Space: Paper's Data** | | | | |
| 2 | Paper ID | PQ1 (Venue Classification) | PQ1 (Venue Name) | PQ2 (Year) | PQ3 (Citations) | |
| 3 | SP1 | | | | | |
| 4 | SP2 | Conference | | | | |
| | | Journal | | | | |
| 5 | SP3 | Workshop | | | | |
| 6 | ... | | | | | |
| 7 | | | | | | |

Figure 3  Publication Space questions related to the paper

The DEF portion related to Publication space for PQ4 and PQ5 is show in the next figure:

|   | A | B | C | D | E | F |
|---|---|---|---|---|---|---|
| 1 | | **Publication space: Author's Data** | | | | |
| 2 | PaperID | Author | PQ4 (number of works) | PQ5 (affiliation) | jbarros: | |
| 3 | SP1 | First author | value | A/I/B | (A)cademic | |
| 4 | SP1 | Second author | value | A/I/B | (I)ndustry | |
| 5 | ... | | | | (B)oth | |
| 6 | SP1 | Last author | value | A/I/B | | |
| 7 | SP2 | First author | value | A/I/B | | |
| 8 | SP2 | Second author | value | A/I/B | | |
| 9 | ... | | | | | |
| 10 | SP2 | Last author | value | A/I/B | | |
| 11 | ... | ... | ... | ... | | |
| 12 | | | | | | |

Figure 4  Publication Space questions related to the Authors

The following table details the data to be extracted for each RQ, the most frequent location of the data in a scientific paper and the proposed classification to be used by our SMS.

*Table 17 Classification of the extracted data for RQs*

| Research Question | Data to be extracted | Probable location | Classification |
|---|---|---|---|
| RQ1: Which patterns? | The name of all the patterns mentioned in the study. | Abstract, Introduction, Content. | Not applicable. |
| RQ2: Which RE activities? | The name of all the activities, tasks, or processes mentioned in the study (pertaining to the requirements engineering phase). | Abstract, Introduction, Content. | Standard for Systems and software engineering — Life cycle processes — Requirements engineering (ISO/29148, 2011), consider all the seven activities (from SR2.1 to SR2.7) and:<br>+ Other (not in the standard)<br>+ Not reported |
| RQ3: impact on what? | The name of the characteristics (of the process or the product), impacted by the use of the patterns. | Abstract, Content, Conclusions. | Open list. Some frequently mentioned characteristics are productivity or quality. When possible use standard ISO 25010 (for Product Quality). |
| RQ4: Was the impact measured? | The metrics and the amount. | Abstract, Content, Conclusions. | Not applicable. |
| RQ5: Which research type is reported? | Research type used. | Abstract, Content. | The classification proposed by Petersen (Petersen, et al., 2015) will be used. (See Figure 5). |

The next figure shows the research type classification as proposed by (Petersen, et al., 2015).

Research type classification (T = True, F = False, • = irrelevant or not applicable, R1–R6 refer to rules).

| | R1 | R2 | R3 | R4 | R5 | R6 |
|---|---|---|---|---|---|---|
| **Conditions** | | | | | | |
| Used in practice | T | • | T | F | F | F |
| Novel solution | • | T | F | • | F | F |
| Empirical evaluation | T | F | F | T | F | F |
| Conceptual framework | • | • | • | • | T | F |
| Opinion about something | F | F | F | F | F | T |
| Authors' experience | • | • | T | • | F | F |
| **Decisions** | | | | | | |
| Evaluation research | ✓ | • | • | • | • | • |
| Solution proposal | • | ✓ | • | • | • | • |
| Validation research | • | • | • | ✓ | • | • |
| Philosophical papers | • | • | • | • | ✓ | • |
| Opinion papers | • | • | • | • | • | ✓ |
| Experience papers | • | • | ✓ | • | • | • |

*Figure 5  Research type (from (Petersen, et al., 2015)*

Some guidelines for classifying works as Evaluation or Validation research type are offered in (Petersen, et al., 2015) and show in the next figure.

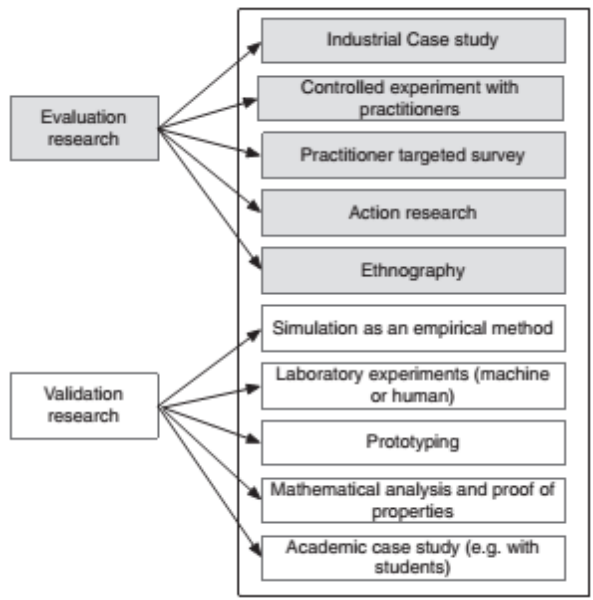

*Figure 6  Several research methods and their respective research type*

*Table 18  Classification of the extracted data for PQs*

| Publication Questions | Data to be extracted | Classification |
|---|---|---|
| PQ1: Venue | Type and name of the venue | For type: Conference, Workshop, Journal. |
| PQ2: Year | Year of publication | |
| PQ3: Number of citations | Quantity of citations to the paper | Reported by Google Scholar[9] |
| PQ4: Productivity of authors | Number of works of each author in the set of selected primary works. | |
| PQ5: Author's affiliation | Affiliation of the authors. | Academy or Industry or Both (A/I/B) |

A value of "Not Reported" is added to any data of the above if the information cannot be extracted from the source.

## 2.5. Data analysis protocol

A detailed guide on how to perform the data analysis and the discussion activity follows:

We first split the set of selected primary works in two halves:

1. For RQs:
    1.1. First half: r3, r5
    1.2. Second half: r1, r6
2. For PQs:
    2.1. First half: r2
    2.2. Second half: r4

To facilitate the writing of this section we suggest producing a paragraph for each RQ and PQ and, discuss/interpret the data presented in the previous section (Results). We will aso: add some mixing graphs (while Results section only has graphs for one item (the result) here we could make comparisons between two or more variables). Data for these graphs come, directly, from columns in the spreadsheet (DEF) and their combinations.

---

[9] Because its data is more updated that data from other sources

Some examples of potential graphs and comparisons to add value to this section include:

- One variable graphs:
    o Patterns (from RQ1): frequency and evolution in time
    o RE activities (from RQ2): (frequency and evolution in time)
    o Impacted feature (from RQ3): classify impact on process or in product
- Two variables:
    o Patterns (from RQ1) crossed with RQ2 (RE activities)
    o Patterns (from RQ1) crossed with RQ3 (Impact)
    o How much was the impact? (from RQ4) crossed with RQ1 (used patterns)
    o How much was the impact? (from RQ4) crossed with RQ5 (Industry's domain)
    o How much was the impact? (from RQ4) crossed with RQ6 (project's domain)

To organize data related to questions RQ1, RQ2 and RQ3 we elaborate the following table:

*Table 19  Combined data fields for RQ1, RQ2 and RQ3*

| Pattern name | RE activity | Feature impacted | Metric and amount | References |
|---|---|---|---|---|
| Pattern name 1 | Activity | Feature | | |
| Pattern name 2 | Activity | Feature | | |
| … | … | … | … | … |

As an example, the analysis of RQ2 and RQ3 can be done with the help of the following graphs:

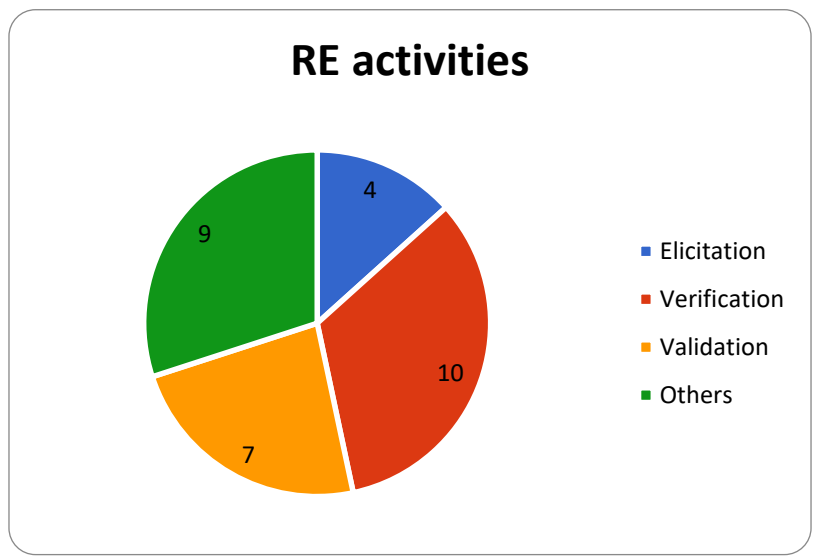

*Figure 7   A pie chart with the number of patterns used in every RE activity.*

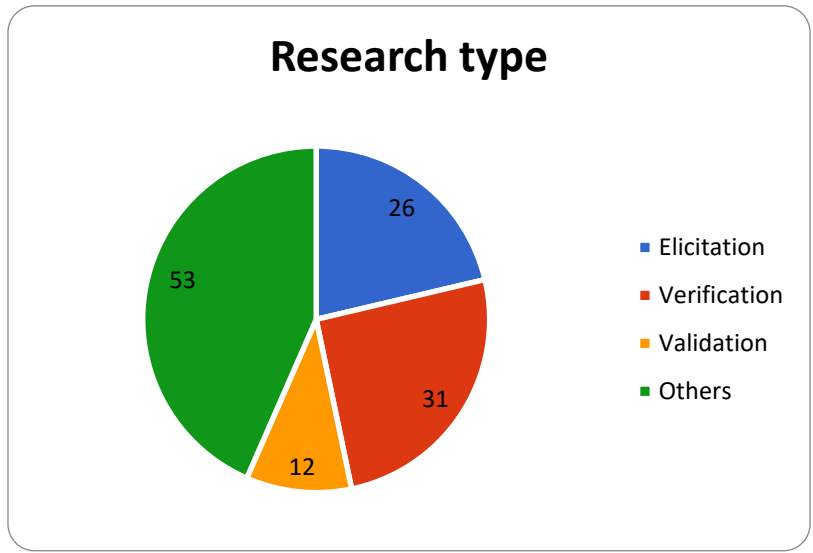

*Figure 8   A pie chart showing the number of works for each research type*

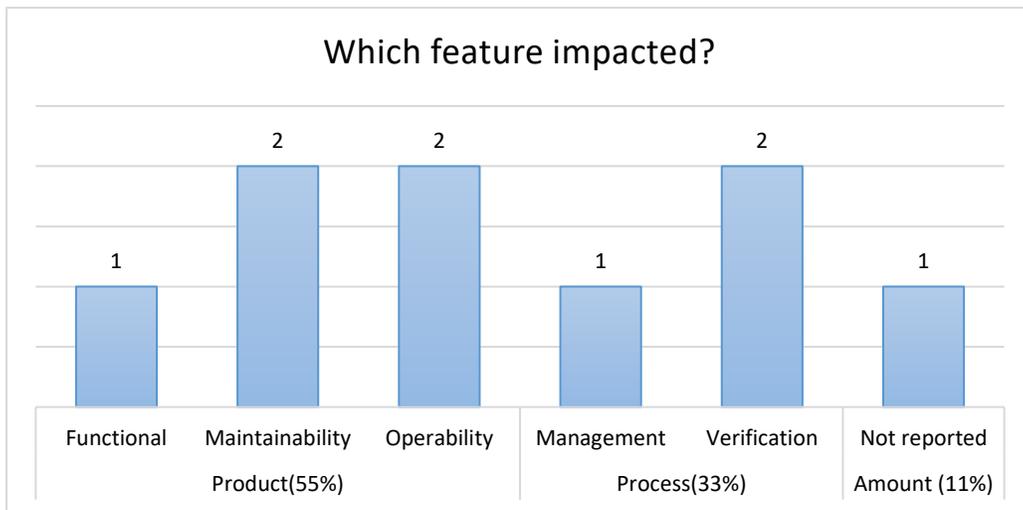

*Figure 9   A bar graph with the total of patterns by classification (process or product) and the impacted feature*

For the analysis of RQ5, we can elaborate a Table relating the features impacted and the research type or a Bubble chart showing the same information.

*Table 20  Features influenced and research type*

| Classification of the Feature | Feature impacted | Research Type | References |
|---|---|---|---|
| Product | Quality | Solution proposal | |
| | Quality | Evaluation research | |
| | Productivity | Opinion paper | |
| | Functionality | Experience paper | |
| Process | Verification | Opinion paper | |
| | Specification | Solution proposal | |
| | Maintenance | Validation research | |

For questions in the Publication space, we can construct a bubble chart to represent the Journals, Conferences or Workshops (venue) and their evolution (published works/year) (PQ1) and Publications per year (PQ2).

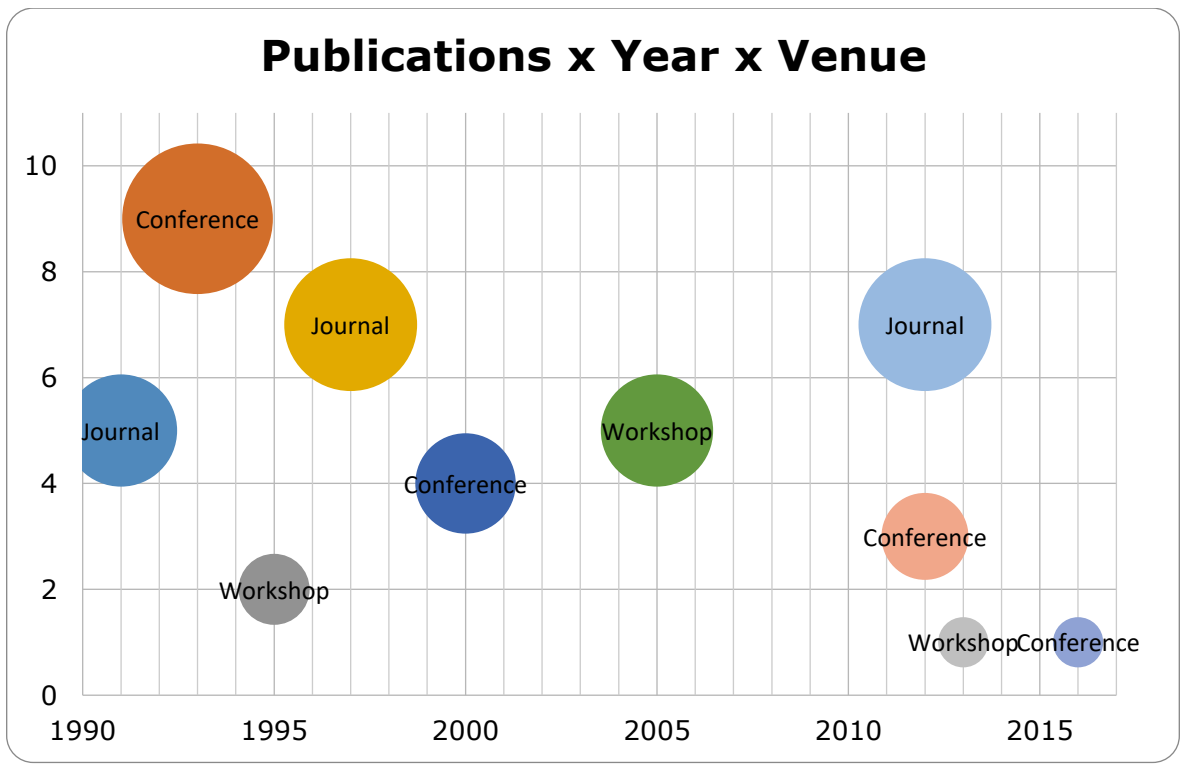

*Figure 10 Number of publications by year and type of venue*

To analyze PQ3 and PQ4 we can construct a table with the references of the selected articles in descending order of the quantity of citations and its classification (A/I/B).

Table 21  Quantity of Citations by Affiliation

| Reference | Quantity of citations to the paper (Google Scholar) | Affiliation (Academy or Industry or Both (A/I/B)) |
|---|---|---|
| Ref#1 | 23 | A |
| Ref#2 | 20 | B |
| … | 15 | I |

In relation to PQ4 we can list the authors considering the number of works of each author in the set of selected primary works.

Table 22  Author's productivity

| Author | number of papers | References |
|---|---|---|
| Name | 3 | ref#1<br>ref#2<br>ref#3 |
| Name | 2 | ref#1<br>ref#4 |
| … | | |

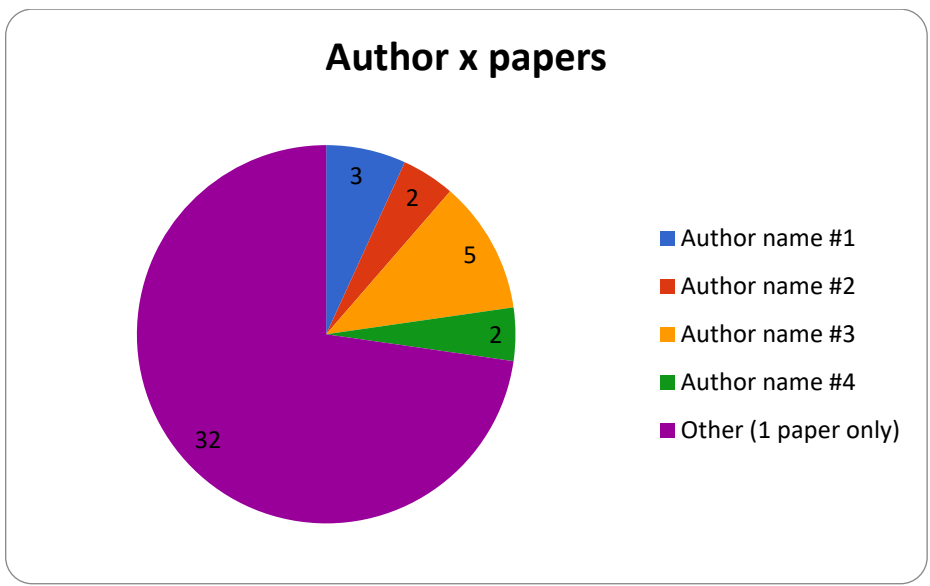

Figure 11 Author's productivity within the set of selected primary papers

### 2.6. Threats to Validity

We will consider the most frequent threats to validity mentioned in other systematic reviews. These are:

- Descriptive validity (the threats associated with the description of data and observations we have made): To reduce it we design a DEF (see Figure 1), objectifying the data extraction process. In addition, reviewers can access the form and revise it; hence, this threat is considered as being under control.
- Theoretical validity (consider two activities):
    - Study identification/selection (missed studies): to reduce this threat we planned three complementary search strategies, with backward and forward snowballing, also checked the DBLP for most cited authors and included the primary works selected by existent SMS and SLR found in our related work section.
    - Data extraction and classification (researcher bias): To decrease the impact of potential researcher's bias we divided the set of primary papers into halves and, assign every half to different researchers who perform the extraction process independently. We will do a double check on every selected paper and for every extracted data item.
    - The Cohen's kappa will be computed for every pair of researchers on every activity mentioned (identification/selection and data extraction).
- Interpretive validity (conclusions drawn are reasonable given the data): we apply the same strategy as for data extraction (divide the set of works between the team of researchers).
- Repeatability (the possibility of achieving the same results by repeating the same processes): in this document, we offer detailed reporting of the research process plus all the data necessary to allow other researchers to repeat our study. In addition, a complete pack of documents and data will be available online (selected studies references, data extraction form, exclusion criteria, graphs, tables and so on).

## Conclusions

We have strictly followed the guidelines proposed by Petersen (Petersen, et al., 2015) to develop this protocol for the SMS. As the whole team adhered to these guidelines to build up the protocol presented in this document, we think the conducting phase of our study could be repeatable without major efforts, and that frequent threats to validity has been mitigated as much as was possible.

# Annex I. Conferences and Journals of interest

The list of potential Conferences to consider for this mapping study include:

**RE**. Requirement Engineering Conference. http://requirements-engineering.org/
**RePa.** International Workshop on Requirements Patterns.
http://www.utdallas.edu/~supakkul/repa16/cfp.html
**ICSE.** International Conference on Software Engineering. http://icse2017.gatech.edu/
**PLoP**. THE INTERNATIONAL CONFERENCE ON PATTERN LANGUAGES OF PROGRAMS.
http://www.hillside.net/plop/2016/
**EUROPLoP**. European Conference on Pattern Languages of Programs. http://europlop.net/
**PATTERNS.** International Conferences on Pervasive Patterns and Applications.
https://www.iaria.org/conferences2016/PATTERNS16.html
**ICPRAM.** International Conference on Patterns Recognition Applications and Methods.
http://www.icpram.org/
**ICPR.** International Conference on Patterns Recognition. http://www.icpr2016.org/site/
**REFSQ.** Conference on Requirements Engineering: Foundation for Software Quality.
https://refsq.org/2017/welcome/

Potential Journals are:

**Requirements Engineering.** http://link.springer.com/journal/766
**IEEE Transactions on Software Engineering.**
http://ieeexplore.ieee.org/xpl/RecentIssue.jsp?punumber=32
**SEKE.** International Journal of Software Engineering and Knowledge Engineering.
http://www.worldscientific.com/worldscinet/ijseke
**TPLOP.** LNCS Transactions on Pattern Languages of Programming.
http://www.springer.com/computer/lncs/transactions+plop?SGWID=0-159502-0-0-0
**SOSYM.** International Journal on Software and Systems Modeling (SoSyM).
http://www.sosym.org/
**JSS.** Journal of Systems and Software.
http://www.journals.elsevier.com/journal-of-systems-and-software
**IST.** Information and Software Technology.
http://www.journals.elsevier.com/information-and-software-technology/

# Annex II. Assessment of the Systematic Mapping

When conducting the study, we will include an evaluation of the work done to "systematically develop" the SMS. It can be used as a self-evaluation, to help authors to check if everything has been done in the right way. Table 23 summarizes all the possible activities to consider when conducting a SMS in a rigorous way.

*Table 23  Identified activities for conducting a Systematic Mapping Study. Adapted from (Petersen et al., 2015)*

| Phase | | Actions | Applied |
|---|---|---|---|
| Need for map | | Motivate the need and relevance | √ |
| | | Define objectives and questions | √ |
| | | Consult with target audience to define questions | --- |
| Study Identification | | | |
| | Choosing search strategy | Automatic search (databases) | √ |
| | | Snowballing | √ |
| | | Manual (Conferences, Main Authors) | √ |
| | Develop the search | PICO | √ |
| | | Consult librarians or experts | √ |
| | | Iteratively try finding more relevant papers | √ |
| | | Keywords from known papers | √ |
| | | Use standards, encyclopedias, and thesaurus | √ |
| | Evaluate the search | Test-set of known papers | √ |
| | | Expert evaluates result | √ |
| | | Search web-pages of key authors | √ |
| | Inclusion/Exclusion | Identify objective criteria for decision | √ |
| | | Add additional reviewer, resolve disagreements | --- |
| | | Decision rules | √ |
| Data extraction and Classification | | Identify objective criteria for decision | √ |
| | | Obscuring information that could bias | --- |
| | | Add additional reviewer, resolve disagreements | --- |
| | | Test–retest | --- |
| | | Classification scheme | √ |
| | | Research type | √ |
| | | Research method | √ |
| | | Venue type | √ |
| Validity discussion | | Validity discussion/limitations provided | √ |

We will also apply the evaluation rubric suggested by Petersen (Petersen et al., 2015) to evaluate our work in terms of all the key activities a SMS should include. The following tables show the rubric criteria. The scores identified by our mapping study will be highlighted (bold text):

Table 24  Rubric: need for review.

| Evaluation | Description | Score |
| --- | --- | --- |
| No description | The study is not motivated and the goal is not stated | 0 |
| Partial evaluation | Motivations and questions are provided | 1 |
| Full evaluation | Motivations and questions are provided, and have been defined in correspondence with target audience | 2 |

Table 25  Rubric: choosing the search strategy.

| Evaluation | Description | Score |
| --- | --- | --- |
| No description | Only one type of search has been conducted | 0 |
| Minimal evaluation | Two search strategies have been used | 1 |
| Full evaluation | Three or more search strategies have been used | 2 |

Table 26  Rubric: evaluation of the search.

| Evaluation | Description | Score |
| --- | --- | --- |
| No description | No actions have been reported to improve the reliability of the search and inclusion/exclusion | 0 |
| Minimal evaluation | At least one action has been taken to improve the reliability of the search OR the reliability of the inclusion/exclusion | 1 |
| Partial evaluation | At least one action has been taken to improve the reliability of the search AND the inclusion/exclusion | 2 |
| Full evaluation | All actions identified have been taken | 3 |

Table 27  Rubric: extraction and classification.

| Evaluation | Description | Score |
| --- | --- | --- |
| No description | No actions have been reported to improve on the extraction process or enable comparability between studies through the use of existing classifications | 0 |
| Minimal evaluation | At least one action has been taken to increase the reliability of the extraction process | 1 |
| Partial evaluation | At least one action has been taken to increase the reliability of the extraction process, and research type and method have been classified. | 2 |
| Full evaluation | All actions identified have been taken | 3 |

Table 28  Rubric: study validity.

| Evaluation | Description | Score |
| --- | --- | --- |
| No description | No threats or limitations are described | 0 |
| Full evaluation | Threats and limitations are described | 1 |